# Massive damage generation accompanying pyramidal slip in hexagonal-close packed magnesium: an origin for its high hardening and brittleness

Yizhe Tang[1,2*]

[1]Shanghai Institute of Applied Mathematics and Mechanics, School of Mechanics and Engineering Science, Shanghai University, Shanghai 200444, China

[2]Shanghai Key Laboratory of Mechanics in Energy Engineering, Shanghai 200444, China

E-mail: [*] yzhe.tang@gmail.com

## Summary

Mg represents a group of technically important hexagonal-close packed (hcp) metals whose mechanical behaviors are very different from body-centered cubic (bcc) and face-centered cubic (fcc) metals and the underlying mechanisms remain poorly-understood. It has been long known that Mg has high hardening and low ductility, and this was conventionally attributed to the scarce of active pyramidal slip. Recently immobilization of pyramidal dislocations was proposed to be the origins. Here we present a different scenario that has hitherto never been reported for Mg: mobile screw pyramidal dislocations double cross-slip and continue to double cross-slip as they glide, generating abundant vacancy clusters on the slipped planes. These vacancy clusters hinder subsequent dislocation motion and thus directly result in hardening; their accumulation as plastic deformation proceeds eventually causes premature fracture, i.e. brittleness. The massive damage accompanying pyramidal slip discovered here provides a different explanation for the long known hardening and brittleness of Mg, and suggests that sufficient amount of pyramidal dislocations could still be insufficient to effectively accommodate plasticity. Stabilizing pyramidal dislocation motion to avoid cross-slip and subsequent damage generation is also essential for property-improvement. The instability of pyramidal slip, manifested as continuous cross-slip here (could be different elsewhere), originates from its huge Burgers vector and could be a common feature of its kind. The detailed cross-slip mechanism also provides insights into the poorly known instabilities and complicated behavior of dislocations in a broader range of crystals with low symmetry and/or large Burgers vectors, including other hcp metals and ceramics.

**Introduction**

Hexagonal-close packed (hcp) metals such as Mg, Ti, Co, Zn and others are of great technical importance in various applications. For example, Mg is the lightest structural metal, 78% lighter than steel; and Ti is the most important light-weight and high-temperature metal in aerospace industry. The hcp lattice structure has relatively lower symmetry and higher anisotropy compared to body-centered cubic (bcc) and face-centered cubic (fcc) structure. As a result, mechanical behaviors of hcp metals are also quite different from common bcc and fcc metals.

Mg is long known for its low ductility and high hardening [1-3]. There was also anomalous temperature dependence reported [1,2,4]. The hcp structure of Mg has a relatively larger number of distinct slip systems that are responsible for its mechanical behavior, including <a> dislocations on basal, prism and pyramidal planes, <c> dislocations on prism plane, and <c+a> dislocations on prism and pyramidal planes [5]. Recently the author and co-workers also reported <c+a> dislocations on a novel {-12-11} plane in Mg [6, 7]. Amongst these possible systems, basal <a> slip is the easiest; yet pyramidal <c+a> slip is a geometrical requisite for *c*-axial normal deformation, since <a> slip has no contribution to *c*-axis due their orthogonality and prism <c> slip can only contribute to shear. In practice, polycrystalline Mg has a strong texturing along *c*-axis, and thus *c*-axial normal deformation during processing, such as rolling, or at service, would inevitably involve pyramidal <c+a> slip. However, the requisite pyramidal <c+a> slip is a difficult mode that would be much less active than basal slip, and the scarce of active pyramidal slip and dominance of basal slip would lead to deformation inhomogeneity and localization, which are conventional causes for hardening and brittleness. Hence, pyramidal slip is accepted to play a key role in Mg's mechanical behavior, and enhanced pyramidal slip was indeed reported to result in improved ductility in Mg-Y alloys with yttrium (Y) addition [8,9] or in confined volumes [10].

Despite its great importance, pyramidal <c+a> slip in Mg remains largely mysterious, mainly due to its large Burgers vector, almost twice of its nearest neighbor distance, and the resultant complicated slip behavior. The existence of <c+a> dislocations in Mg was first confirmed by postmortem Transmission Electron Microscopy (TEM) examinations back in 1970s [1,2]. The habit plane was reported to be the {-12-12} second-order pyramidal planes (Py-II) [2]. The viewpoint of Py-II <c+a> slip in Mg predominated therefrom, in spite of claims on the impossibility of full <c+a> dislocations with such a large Burgers vector [11].

The fundamental processes of pyramidal slip and relevant phenomena typically occur at the atomic scale within nanoseconds, therefore direct experimental studies of pyramidal dislocations, especially their motion using techniques such as in situ TEM are still challenging; while time-dependent density functional theory (DFT) calculations of dislocation motion are also computationally too expensive, if not impossible. Atomistic simulations, on the other hand, feature excellent balance between computational accuracy and efficiency, and thus play an essential and unique role in these quests. In 2014, the author and co-worker demonstrated that <c+a> dislocations form on the {1011} first-order pyramidal planes (Py-I) planes first, and then transform onto P-II planes, suggesting that co-existence would be a more plausible scenario [12]. Later on, Xie et al [13] reported that <c+a> dislocations were actually on Py-I planes, not Py-II. Recent in situ TEM study of *c*-axis compression indeed showed co-existence of Py-I and Py-II <c+a> dislocations [10].

In addition to the mystery of habit planes, pyramidal <c+a> dislocations in Mg also have exhibited features distinct from dislocations in common bcc and fcc metals. For example, atomistic simulations predicted significant asymmetry when the dislocation is subjected to opposite shear stresses. The asymmetry exists not only in the Peierls stresses [6,14], but also in their slip behaviors [6,15]. Active cross slip of a Py-II screw dislocation to prism and a novel {-12-11} planes [6] was also reported whereas gliding of Py-II screw dislocation on its original habit plane has not been predicted by atomistic simulations so far. In addition to the asymmetry, other unusual slip behaviors were also reported, including shuffling and climb during gliding of a near-edge dislocation on Py-I plane [12, 14], and shuffling and formation of superjogs during gliding of a screw dislocation on Py-II plane [16].

Recently, it was reported that the edge <c+a> dislocations on Py-II planes tend to transform into sessile cores, and the immobilization of the edge <c+a> dislocations was believed to be the origins of the hardening and brittleness in Mg [17]. The newly proposed scenario of dislocation immobilization is totally different from the conventional one concerning scarce of pyramidal slip, and provides new insights into the complicated and mysterious behavior of pyramidal dislocations. It was also reported that enhanced ductility can be achieved by modulating cross-slip of <c+a> dislocations through specific alloying [18] and application of non-glide stress [19].

The present study continues the author and coworker's previous work in quest of fundamental understanding of mysterious pyramidal slip in hcp Mg [12], with focus on <c+a> dislocations on Py-II planes [20]. We report that, massive vacancy clusters (permanent damage at the atomic scale) are generated and left behind as glissile screw <c+a> dislocations glide on the Py-II planes, directly leading to hardening and brittleness of Mg. The massive damage generation arises through a continuous double cross-slip mechanism of screw dislocation in motion, suggesting intrinsic instability of pyramidal slip. The causes of this specific insatiability of pyramidal slip are also revealed through the energy landscapes associated with dislocation motion, which provide general guide for stabilizing slip pyramidal and improving properties.

**Defect evolution during *c*-axis compression: dominance of screw *<c+a>* slip**

The presence of dislocations of interest is ensured by compressing a pristine single Mg crystal along *c*-axis, a condition that maximally facilitates pyramidal slip and avoids basal slip and extension twins. The computational details are the same as that in the previous work [12]: 50×50×50 *nm* sample loaded at a strain rate of $10^8$ $s^{-1}$ from 0 *K* using LAMMPS code [21], and VMD for visualization [22] and CNA for defect detection [23]. The only difference is the adoption of a different potential, a revised MEAM potential [24, 25] that predicts better, planar <c+a> dislocation cores on Py-II planes [20]. Larger sample size of 100 *nm*, higher strain rate of $10^9$ $s^{-1}$ and room temperature of 300 *K* are also adopted and all show similar results, and thus are not shown here.

The sample experiences elastic deformation first and plasticity initiates at a relatively high stress of ~4.1 GPa as defects start to nucleate from sample corners. The high stress is required here to nucleate defects, due to the intactness of the sample, rather than drive pre-existing defects to propagate. The corresponding resolved shear stresses for <c+a> slip on Py-I and Py-II planes are 1.64 and 1.83 GPa, respectively. The nucleation of defects in the incipient stage of plasticity is expectedly the same as that in our previous work [12], despite the difference in the potentials adopted. As shown in Fig. 1, at a strain of $\varepsilon$=6% (time reset to 0 hereafter), stacking faults (SFs) start to nucleate on Py-I planes first, as seen from Py-I planes' flatter feature, from the upper-left and lower-left corners. The cooperative slip of SFs on multiple Py-I planes then lead to formation of <c+a> dislocations on Py-II planes, indicated by two red dash lines at $t$=5 *ps*. As full <c+a> dislocations on Py-I planes form and glide, their screw components start to cross slip to Py-II planes, as evidenced by the connecting of Py-I (green dash lines) and Py-II (red dash-dot lines) dislocation segments in the lower-left corner at $t$=5 *ps*, and the upper-left corner at $t$=10 *ps*. At $t$=10 *ps*, most of the dislocations are on Py-II planes. At $t$=15 *ps*, most of the two Py-II dislocations formed through cooperative slip, indicated by red dash lines, have aligned to their Burgers vectors ***b***=<***c+a***> (see yellow arrows representing ***b***) to become screw-like.

The dominance of screw <c+a> dislocations at later stage of plasticity is due to the low Peierls stress and high mobility of the edge components. Peierls stress evaluations of an infinite Py-II <c+a> screw and edge dislocation using the same method as that in our previous work [6] were performed. The samples have dimensions of 60×60 *nm* along the normal direction of slip plane and the gliding directions, and 2*b* along dislocation line direction with periodicity. The cores were initially introduced as two 1/2 <c+a> partials separated by a distance of about 2 nm, according to the elastic displacement fields around dislocation, followed by energy minimization. Shear stress was applied by homogeneously straining the sample. Then NVE simulations were performed while fixing the top and bottom 3-layer atoms only in the shear direction. The calculated Peierls stresses for Py-II screw and edge <c+a> dislocation are 300 and 140 MPa, respectively. Hence, if there are no constraints, like periodicity, to force a mixed dislocation to stay edge-like, its line direction will become screw-like as it glides freely, in analogy to the dominance of screw dislocations with higher Peierls stress in bcc metals.

For the two dislocations formed by cross-slip, their screw segments are too close to the sample's left (01-10) surface and are gradually exiting as they develop, leaving mostly edge-like segments within the sample; yet immobilization of these edge segments is not observed. This probably is because the characters of these edge-like segments are not pure edge, and the low temperature (several *K*) and limited time scale (several picoseconds) are not sufficient for the transition to happen.

Despite the higher stress level and smaller sample size, the overall evolution of pyramidal dislocations observed here is largely consistent with that during *c*-axis compression of a sub-micrometer-size Mg sample studied with in situ TEM [10].

**Massive point-defect generation**

The propagation of dominating screw *<c+a>* dislocations on Py-II planes in later stage of plasticity exhibits a feature that is very distinct from common fcc and bcc metals: as the screw dislocations glide forward, abundant point defects, specifically vacancy clusters, are generated and left behind on the slipped Py-II planes, as seen at $t$=10 and $t$=15 *ps* in Fig. 1; while formation of superjogs [16] is not evidenced. At $t$=30 *ps*, the two screw dislocations have all exited the sample, and the Py-II planes they ever glided on are full of vacancy clusters, as if they were subjected to intense irradiation and were severely damaged internally. Contrastively, much less point defects are observed during slip of the other two dislocations with mostly edge segments. Such feature accompanying screw dislocation slip is actually very unusual. Typically, simple slip of dislocations proceeds without generating any extra defects, since the Burgers vector is usually the minimum period to assure translation invariance, and the lattice retrieves intactness after slip. The Burgers vector of *<c+a>* dislocations in hcp lattice is also a period, although not minimum, to assure translation invariance, and thus defects, especially vacancies, are also not expected here in compression. In common fcc and bcc metals, the occasions that dislocation slip is accompanied with extra defect generation (only in very limited amount) usually involve complicated local interactions with other defects, or local diffusion processes prompted at elevated temperatures, which are all irrelevant to the case here.

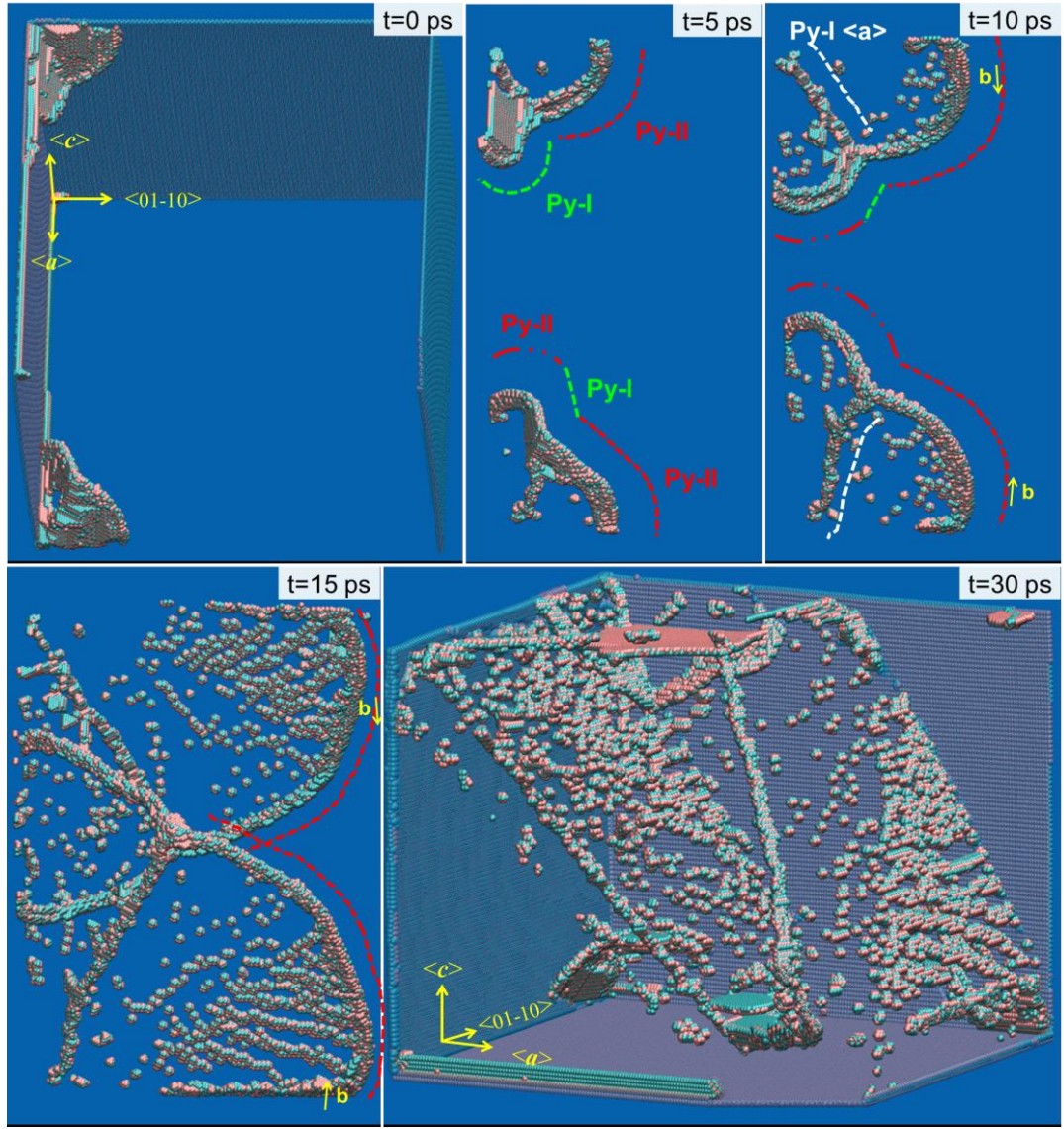

Figure 1 Sequences showing nucleation and propagation of pyramidal $<c+a>$ dislocations during *c*-axial compression of hcp Mg. Only non-hcp atoms are shown, and pink and cyan spheres represent A and B type atoms, respectively, where . . .ABAB. . . is the packing sequence of basal planes in hcp lattice. The time has been reset to 0 at a strain of 6%. $<c+a>$ dislocations on Py-I and Py-II planes are indicated with green and red dash lines, respectively. The Burgers vectors of $<c+a>$ dislocations on Py-II planes are indicated with yellow arrows and $<a>$ dislocation on Py-I planes are indicated with white dash lines. Abundant vacancy clusters, spreading on the Py-II planes that $<c+a>$ dislocations ever glided, are clearly visible after *t*=10 *ps*.

It is worth noting that, this unusual feature is still the same, unexpectedly [20], as that in our previous work [12], despite the adoption of a potential predicting better Py-II dislocation cores here. The commonness of this feature shared by two different types of potentials further justifies its physical plausibility, and hence re-attracts our attention. The focus of the present study thus is placed on how and why this unusual feature forms.

**How point defects are generated: reversal of shift direction**

A careful examination is performed first to reveal how point defects are generated as screw <c+a> dislocation on Py-II plane passes by. A screw dislocation emitted from the upper-left corner, indicated by a red dash line in the upper half of Fig. 1, $t=5$ *ps*, is selected as a representative dislocation. When the dislocation is well developed at a later stage of $t=15$ *ps*, its slip plane, defined as a virtual Py-II plane across which atoms undergo opposite shifts of $\pm \boldsymbol{b}/2$, is identified. According to the location of the slip plane, the atomic layer of Py-II atoms that were originally right below the slip plane (referred to as "lower layer" hereafter ) before dislocation's presence (at $\varepsilon=0\%$) can be backtracked, as seen in Fig. 2, $t=15$ *ps*. By tracking the movements of atoms in the lower layer, patterns of pyramidal slip and associated phenomenon can be clearly revealed. For better visualization, two arrays of atoms in a direction [01-10] perpendicular to screw dislocation line (or $\boldsymbol{b}$) in the lower layer are highlighted with larger spheres.

As seen in Fig. 2, at $t=0$ *ps*, in the absence of dislocation, there is no visible shifts detected. At $t=11$ *ps*, the screw dislocation, indicated by red dash line, enters from the left side and moves towards right. A clear upward shift of $+\boldsymbol{b}/2$, a distance between two neighboring [01-10] arrays, is visible for atoms on the left, slipped side of the dislocation; while atoms on the right, fresh side remain un-shifted. For the atomic layer above the slip plane (referred to as "upper layer" hereafter), atoms on the left, slipped side undergo an opposite shift, downwards by $\boldsymbol{b}/2$. A complete relative shift of $\boldsymbol{b}$ across the slip plane is thus fulfilled. As the screw dislocation propagates, peculiar events happen. Taking the highlighted pink array as an example, at $t=16$ *ps*, three atoms in the pink array stop following their predecessors (atoms to their left) to move upwards; instead they stay at their original positions. At *t*=16.85 *ps*, several successors of these three (to their right) start to move in the opposite direction, namely downwards, by *b*/2. As more successors join the process, the original upward shift of +*b*/2 in the pink array has been reversed to downward, as seen in Fig. 2 at *t*=18.6 and *t*=20.2 *ps*. After the dislocation exits the sample at *t*=24.75 *ps*, the atoms with upward shift on the right side of the pink array clearly exhibit a visible relative shift of *b* to those with downward shift on the left side; while the three atoms in the middle, although seem like shuffled as they blend in with a cyan array, actually experience no net shifts and are merely part of the reversal process. Such a sudden reversal of shift direction during dislocation slip is actually very unusual.

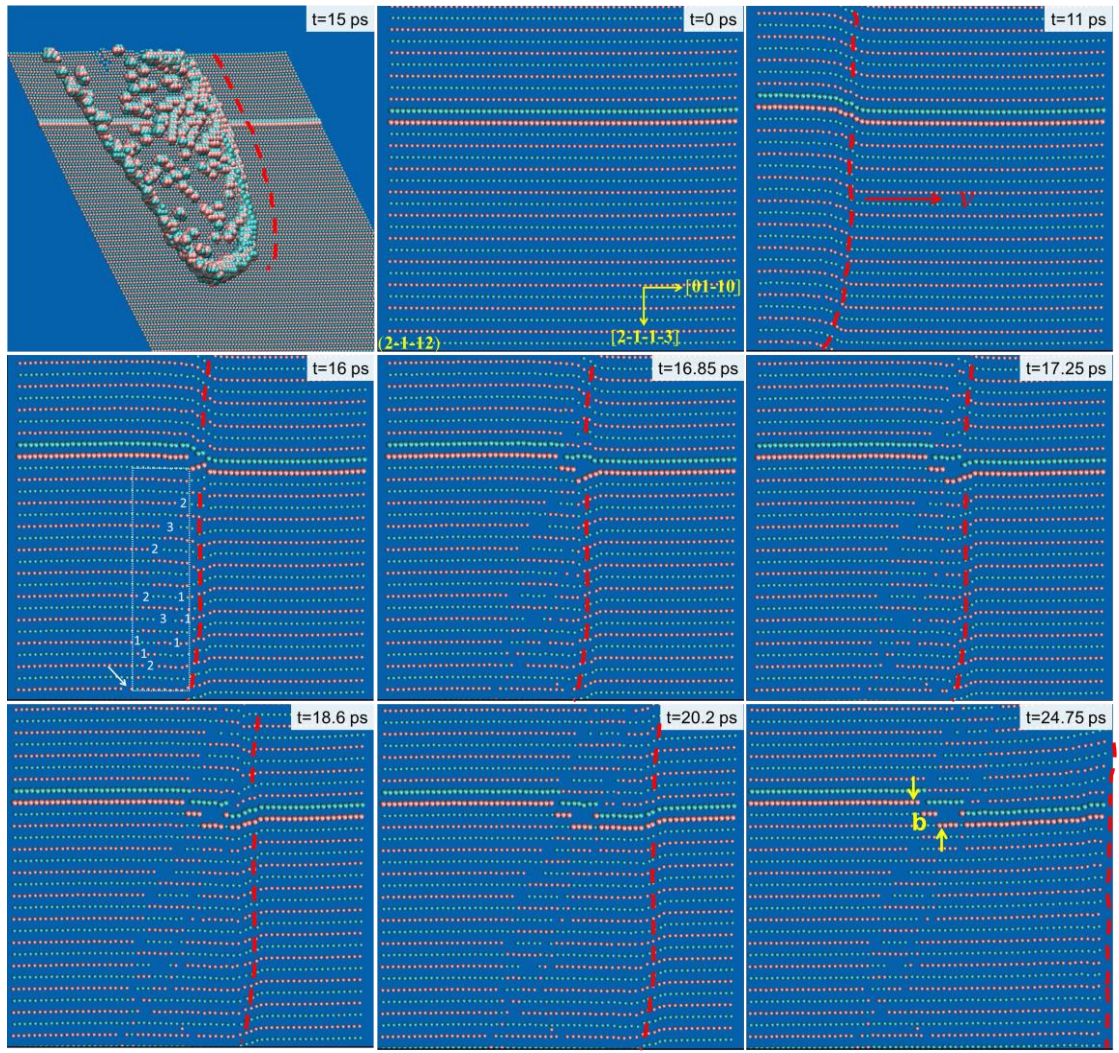

Figure 2 Sequences showing atomic movements in the lower layer of a representative Py-II screw <*c+a*> dislocation. Two arrays of atoms in a direction [01-10] are highlighted with larger spheres

and the dislocation line is indicated with red dash line. The relative positions of the lower plane, the dislocation and the two arrays are shown at $t$=15 *ps*. Reversal of the upward shift of *b*/2 on the left, slipped side of the dislocation to downward shift starts at t=16.85 *ps* and t=17.25 *ps* for the highlighted pink and cyan array, respectively, giving rise to the formation of a three-vacancy cluster in between.

Similar reversal happens to the cyan array as well, starting at ~17.25 *ps*, slightly later than the pink array. If all arrays experience the reversal simultaneously, no extra defects would arise. However, when the reversal starts in the pink array, the neighboring cyan array still remains a regular upward shift. Opposite shifts of neighboring arrays thus lead to split of the two, creating an empty space in between, as seen in Fig 2 at $t$=16.85 *ps*. The empty space, or vacancy cluster speaking of lattice, has a width equal to two arrays of atoms, and its length grows as the dislocation moves forward. By the time the cyan array joins the reversal later at $t$=17.25 *ps*, a cluster containing three vacancies is already permanently formed.

The peculiar reversal happens not only in the two highlighted arrays; a lot more arrays in the atomic plane experience similar events, as evidenced by the emerged vacancy clusters, seemingly shuffled atoms and relative shifts in the arrays. Without unusual reversal, all the arrays should have retrieved intactness after dislocation passage and appear the same as that before dislocation passage ($t$=24.75 vs. $t$=0 *ps* in Fig. 2). It is the reversal of shift direction during dislocation slip that directly leads to formation of abundant point defects, as observed in Fig. 1.

**Mechanism of point defect generation: double cross-slip and jog movement**

The reversal of shift direction actually causes movement of slip plane of a dislocation. For a screw dislocation, change of slip plane conventionally happens when the screw dislocation cross-slips to another plane. Here the dislocation core dissociates on the Py-II planes and stays on them during the whole process, even though it has moved down/up perpendicular to its slip plane by a layer. If we only judge from the consequences and have to call it cross-slip, then it is actually a double cross-slip process. The screw dislocation cross-slips vertically from Py-II to the prism plane, and then cross-slips vertically back to Py-II plane. However, if we consider the specific processes involved, this double cross-slip process observed here is so peculiar:

a) The core always stays on Py-II planes and never transitions to other planes.

b) The distance the screw has double cross-slipped is only one layer, about 0.136 nm, even less than 1/10 of the size of the dissociated core itself on the Py-II plane (e.g. the distance between the two dissociated partials, at least 1.6 nm).

c) The Schmid factor of any slip in the prism plane is zero during both c-axial normal deformation and pure shear in Py-II planes, and thus there is no resolved shear stress on the prism plane. Hence, the observed cross-slip to prism plane is definitely not driven by shear stresses.

It is indeed an instability event happening in the large <$c+a$> core, but it's nothing like those involved in conventional cross-slip processes for extended dislocations, such as Friedel-Escaig's constriction mechanism [26], Fleischer's folding partial mechanism [27] and surface-induced cross-slip mechanism [28]. On the contrary, the change of slip plane by one layer observed here is more similar to the climb process of an edge dislocation. The only difference is that, climb of an edge dislocation generates vacancies/interstitials, whereas cross-slip of a screw does not. This is because any plane containing the screw line could be a slip plane, and thus transition between them wouldn't generate any defects. Vacancies were generated only when part of the dislocation line cross-slipped and part of it didn't such that a jog was formed and moved (asynchrony of cross-slip).

This climb-like double cross-slip of screw dislocation caused by instability has hitherto never been reported for extended dislocations. The detailed mechanism of this double cross-slip and asynchrony of double cross-slip is schematically demonstrated in Fig. 3. The first row of Fig. 3 shows relative shifts around a screw dislocation, indicated by red dash line. The slip plane is Py-II plane (2-1-12) and is parallel to the plane of paper, the screw dislocation line direction and Burgers vector are along the vertical direction [2-1-1-3], and its gliding direction is along the horizontal [01-10] direction. The left side of the dislocation has upward (downward) shift relative to the right fresh side in the lower (upper) layer, indicated by solid (dash) black lines, in analogy to the scissors. If arrays in the lower half of the lower layer suddenly reverse their shift direction, as indicated by blue lines in the second row of Fig. 3, the lower and upper layers now have the same shift direction, and the lower layer and the layer below it has opposite shift directions. The slip plane, always being defined as the virtual plane across which opposite shifts occur, thus moves down (perpendicular to the plane of paper) one atomic layer for the lower segment of the dislocation. That is to say, the lower dislocation segment double cross-slips down by an atomic layer through the unusual reversal of shift direction. A tiny jog of one atomic layer high with line direction [2-1-12] perpendicular to Py-II plane or plane of paper, connecting the cross-slipped lower and un-cross-slipped upper segments is thus formed, as indicated by green solid line in the side view of Fig. 3, second row.

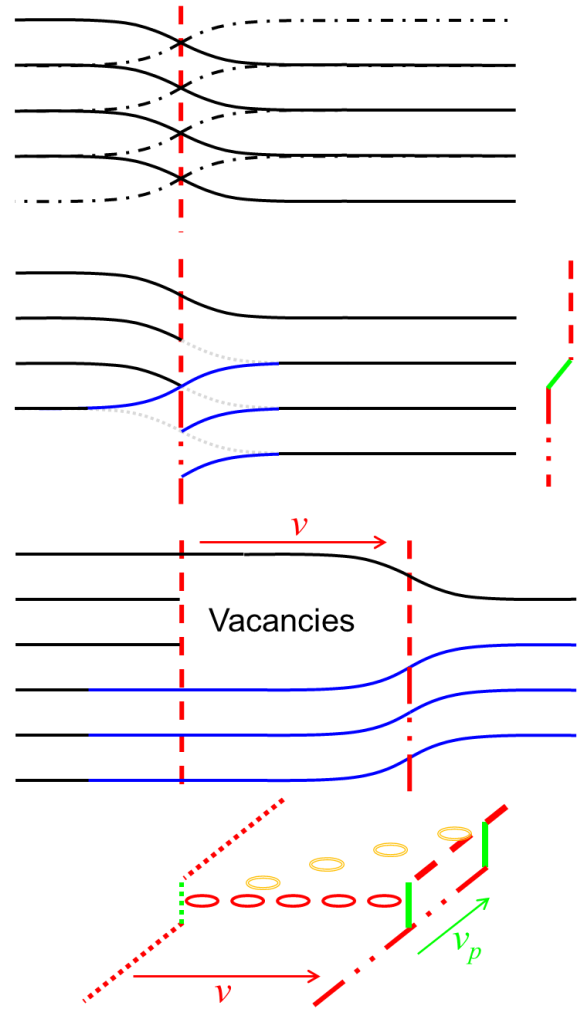

Figure 3 Schematic showing opposite shifts in lower (black solid line) and upper (black dash line) layer around a screw dislocation (first row); reversal of shift direction in the lower layer leading to double cross-slip down of the dislocation's lower segment and formation of a one-layer tiny jog (second row); motion of the jog along with the dislocation at velocity $v$ generating continuous vacancy cluster (third row); motion of the jog along both the dislocation's velocity and line directions generating discrete vacancy clusters (fourth row). The original and double cross-slipped dislocation lines and the jog are indicated by red dash, red dash-dot and green line, respectively. The reversed arrays in the lower layer are indicated by blue lines. The continuous and discrete vacancy clusters are indicated by red and orange ovals, respectively.

Since the screw core dissociates along horizontal [01-10] direction, the jog also has to dissociate horizontally along [01-10], and thus resides on (2-1-13) plane, namely the cross product of [21-12] and [01-10]. Apparently the Burgers vector of the jog [2-1-1-3] happens to be the normal of its residing plane (2-1-13), and thus motion of the jog would definitely generate defects due to its Burgers vector's out-of-plane feature.

As the dislocation moves forward with a velocity $v$, the jog is carried along with it too; split of the two neighboring arrays then generates two arrays of vacancies that stay attached to the jog, as seen in the third and fourth rows of Fig. 3. It is the asynchrony of double cross-slip between different dislocation segments, not double cross-slip itself that leads to formation of jog and vacancies. The length of the vacancy cluster also grows at a velocity $v$ with the jog. In addition to $v$, the jog can also have a velocity $v_p$ parallel to dislocation line. This happens when longer dislocation segment double cross-slips down as more atoms in the upper dislocation segment join the reversal process. As a result, the generated vacancies also spread along $v_p$, appearing as discrete clusters, indicated as orange ovals in the fourth row of Fig. 3. In fact, both continuous arrays and discrete clusters of vacancies are present in Fig. 1. If the entire upper dislocation segment double cross-slips down all of a sudden, the jog then disappears, leaving vacancy clusters that are already formed isolated.

Similarly, if upper half of the lower layer reverses shift direction first, the upper dislocation segment then double cross-slips down, forming a jog with opposite line direction to the one shown in Fig. 3. As the jog moves forward, two arrays of interstitials, instead of vacancies, will be generated. Interstitials, although possible, are rarely present in Fig. 1, implying the preference of vacancies over interstitials. Reversal of shift direction in the upper layer is the same to that in the lower layer, except that the dislocation segment now double cross-slips up, instead of down.

The double cross-slip process revealed here only involves reversal of atoms in one atomic layer, and yet the one-layer double cross-slip can continue to take place at different locations along the dislocation line, generating multiple stair-like tiny jogs in the dislocation line. These one-layer high jogs are too tiny to be distinguished by naked eye in Fig. 1; however, their existence and locations are clearly evidenced by the vacancy arrays formed behind them as they pass by. Double cross-slip, once by a layer, and continuous double cross-slip are thus the primary processes responsible for the unusual massive generation of point defects during slip of Py-II screw dislocation; whereas simultaneous reversal in multiple layers that leads to double cross-slip by multiple layers and formation of a longer jog with multiple-layer height, although possible, is not observed here. Extension of a tiny jog along dislocation motion direction to form a wider jog, caused by one of the connecting dislocation segment moving faster than the other, is also possible but not observed.

Generally, jogs are known to form by dislocation intersection, climb or even complicated cross-slip. Recently, Yi et al. reported cross-slip of a screw <a> dislocation on prismatic plane in Mg, and asynchrony of cross slip between basal and prismatic planes led to jog formation: kinks in the basal plane became jogs in the prismatic plane [29]. In the present study, the jog is formed by asynchrony of double cross-slip; however, the specific mechanism for the cross-slip, namely reversal of shift direction during slip, is totally different from other known cross-slip mechanisms [26-28]. This novel mechanism is essentially an instability event happening at the <c+a> core, and has not been reported. All other phenomena including jog formation by asynchrony of double cross-slip, and vacancy generation by jog motion are only the consequences of the double cross-slipp or instability.

Vacancy generation by climb of an edge dislocation, or by jog motion due to the jog's out-of-plane slip vector is also well known [5]. Yi also reported diffusion-less climb of edge <*a*> dislocation on both prismatic and Py-I plane induced by solute in Mg [30].

The number of vacancies *n* generated by a one-layer high jog traveling a distance *l* along [01-10] can be expressed as $n = 2d \times l$, where $d = (\sqrt{3}\, a)^{-1}$ is the linear density (atoms/m) along the motion direction [01-10] and $a = 3.19$ Å is the lattice constant. A jog moving along dislocation line yields no defects. For example, in Fig. 2 at *t*=16 *ps*, the jog first comes into view at a visible vacancy location, indicated by a white arrow, and moves along both [01-10] and dislocation line directions. By the time the pink array is about to double cross-slip, the jog has traveled along [01-10] by ~ (10±1) ($\sqrt{3}\, a$) and along dislocation line by 20 (*b*/2), and counting of vacancies formed in each array in the area of the jog's passage, indicated by a white rectangular, sums up to 19, close enough to 20±2.

The total number of vacancies *N* generated by motion of multiple jogs can be expressed as

$N = \Sigma\, 2d \times l_i$

where $i=1\sim k$ and *k* is the total number of jogs, $l_i$ the distance the $i^{th}$ jog has traveled. If in Fig. 1 we assume *k*=20, and all $l_i$ equals to half of the sample size 250 Å, then an estimation of $N \approx 1800$ vacancies is obtained out of 6 million atoms.

In short, as schematically demonstrated in Fig. 3, the observed point defect generation can be interpreted from the perspective of conventional dislocation mechanism [5]: the point defects directly arise from the motion of tiny jogs in the screw dislocation line; and the jogs are formed as a result of double cross-slip and continuous double cross-slip of dislocation; only that the double cross-slip occurs, unconventionally, through unusual reversal of shift direction during slip, not by conventional diffusion or dislocation intersections. The unique reversal process revealed here is thus the key to understanding of the unusual slip behavior of Py-II screw dislocations in Mg.

**Why reversal of shift direction happens**

1. **Other explanations?**

Next we turn to the question why the reversal of shift direction, the double cross-slip and jog formation are happening during pyramidal slip. Surely there will always be simple explanations. For example, one might speculate that there might be jogs forming during formation of Py-II dislocations through cooperative slip. It is simply the motion of these pre-existing jogs along dislocation motion direction that generates point defects, and the observed reversal of shift direction is only the propagating of a jog along dislocation line. If that is the case, there will be vacancies formed as long as the screw dislocation moves, which are actually absent at early stage, as seen at $t=5$ *ps* in Fig. 1. Besides, the speculation of pre-existing jogs also couldn't explain why there are so much more new jogs and vacancies formed subsequently.

Another speculation is that, the Py-II dislocation is always connected to another Py-II dislocation and a Py-I <*a*> dislocation [12] at a triple junction in the interior of the sample, and motion of the junction might exert drag on the dislocation and cause continuous emission and propagation of jogs from the junction. If the double cross-slip and jog formation during pyramidal slip are indeed extrinsic and caused by motion of triple junction, then one would not expect them in the absence of junctions. However, when the junctions are removed by cutting out a 10 *nm* slab from the left side of the sample at $t=10$ *ps*, and all Py-II dislocations start and end at surfaces, continuous double cross-slip and abundant vacancy clusters are still present, confirming that the double cross-slip and jog formation are not caused by junction drag.

**2. Polymorphism in equilibrium dislocation core**

In a more general case of infinite long dislocation with no ends, surface controlled/induced cross-slip is also absent, would the double cross-slip still happen intrinsically? And if it does, why is it happening? To address these questions, an infinite Py-II <*c+a*> screw dislocation is reproduced at the center of a pristine sample using the same method as that in our previous work [6]. The sample has dimensions of $60{\times}60$ *nm* along the [2-1-12] and [01-10] directions, and 2*b* along dislocation line direction [-2113] with periodicity.

The cores were initially introduced as two 1/2 <*c+a*> partials separated by a distance of about 2 *nm*, according to the elastic displacement fields around dislocation, followed by energy minimization. Difference in the separation distance would result in different cores.

Examination of the equilibrium configuration, however, first reveals a polymorphism in the dislocation core structure that is uncommon for fcc and bcc metals. Fig. 4 shows the equilibrium atomic configuration in both the lower (left column) and upper (right column) layer of several typical screw cores obtained.

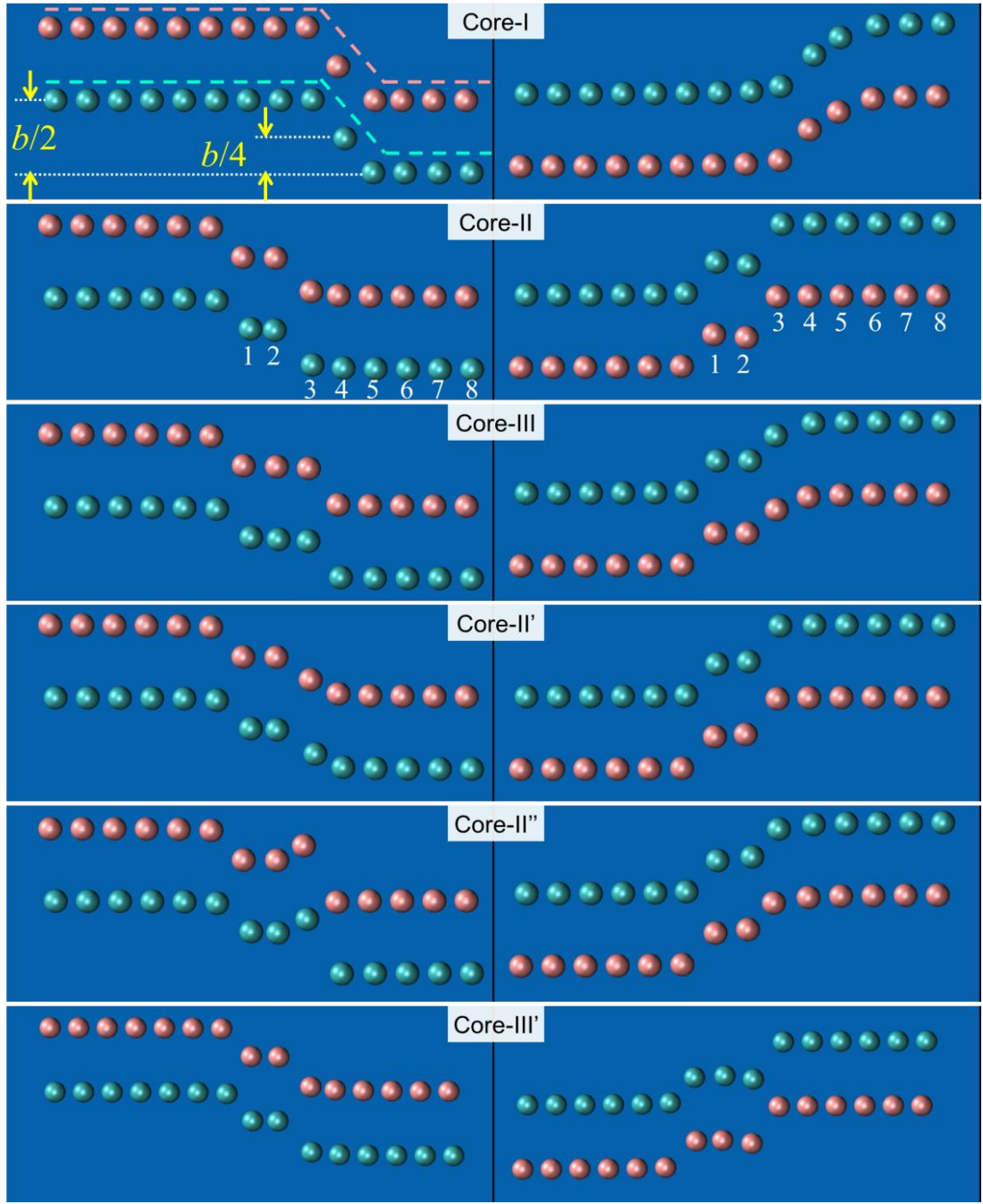

Figure 4 Typical equilibrium configurations in both the lower (left column) and upper (right column) layer showing polymorphism of a Py-II screw <*c+a*> dislocation core. These cores can be simply categorized into three types (I, II and III) by the number of atoms staying at the middle sites (M-sites) of two neighboring arrays with $b/4$ shift. Taking Core-II as reference, the relative energies of Core-I, Core-III, Core-II', Core-II'' and Core-III' are +0.26, -0.030, -0.040, -0.014, -0.034 eV/2$b$, respectively.

When modeling dislocation behavior by introducing <$c+a$> dislocation core, researchers might end up with different core configurations, and thus different results. In order to avoid any possible inconsistencies in dislocation modeling in the future, these cores are presented here to serve as a reference. These cores are similar to those shown in Figs. 2 and 3, all with relative upward and downward shift in the lower and upper layer, respectively. The middle sites of two neighboring arrays in the lower or upper layer, referred to as $M$-sites hereafter, are very special since atoms at these sites have half of the full shift. Then we can simply categorize the cores into three types by the number of atoms staying at $M$-sites. For example, the lower layer in the first, second and third row of Fig. 4 has one, two and three atoms staying at the $M$-sites, and is thus named as "Core-I", "Core-II" and "Core-III", respectively. The lower layer in the fourth and fifth row has two atoms staying at the $M$-sites, and an additional atom with a shift between 0~$b$/4 and $b$/4~$b$/2, and is thus named as "Core-II'" and "Core-II''", respectively. Contrary to "Core-III", in the sixth row there are two atoms staying at the $M$-sites in the lower layer, and three in the upper layer. Actually this core is identical to "Core-III" and is named as "Core-III'" here. Given the large Burgers vector of the <$c+a$> dislocations, the polymorphism in the core structure revealed here is not that surprising.

## 3. Atomic movements during slip

Despite the difference in core structures, the energies and behaviors of these cores turn out to be very similar, except for Core-I. The Core-I, with a higher energy of 0.26 eV/2$b$, is confirmed to be the "Stuck-core" we reported previously [6], which stays stuck at stresses above Peierls stress of 300 MPa and starts moving only when higher stress of ~600 MPs is reached. All the other cores start to move at ~300 MPa and keep moving on Py-II planes until they exit from surfaces. The double cross-slip, however, is difficult to be detected here, simply because the periodicity assures synchronous atomic motion along dislocation line. Even if there is double cross-slip happening, the entire dislocation double cross-slips together and there will be no jogs or point defects formed to reflect the double cross-slip process. Then we have to turn to the atomic movements of atoms around the dislocation core. Taking Core-II as an example, Fig. 5 records the atomic shifts in the core when subjected to a negative shear stress of 470 MPa, ~60% higher than the Peierls stress. Shifts for all atoms in cyan array (motions of pink and cyan arrays are identical due to periodicity) that haven't completed a full shift of $b/2$ in both the lower and upper layer, labeled in order from 1 to 8 in Fig. 4, are shown.

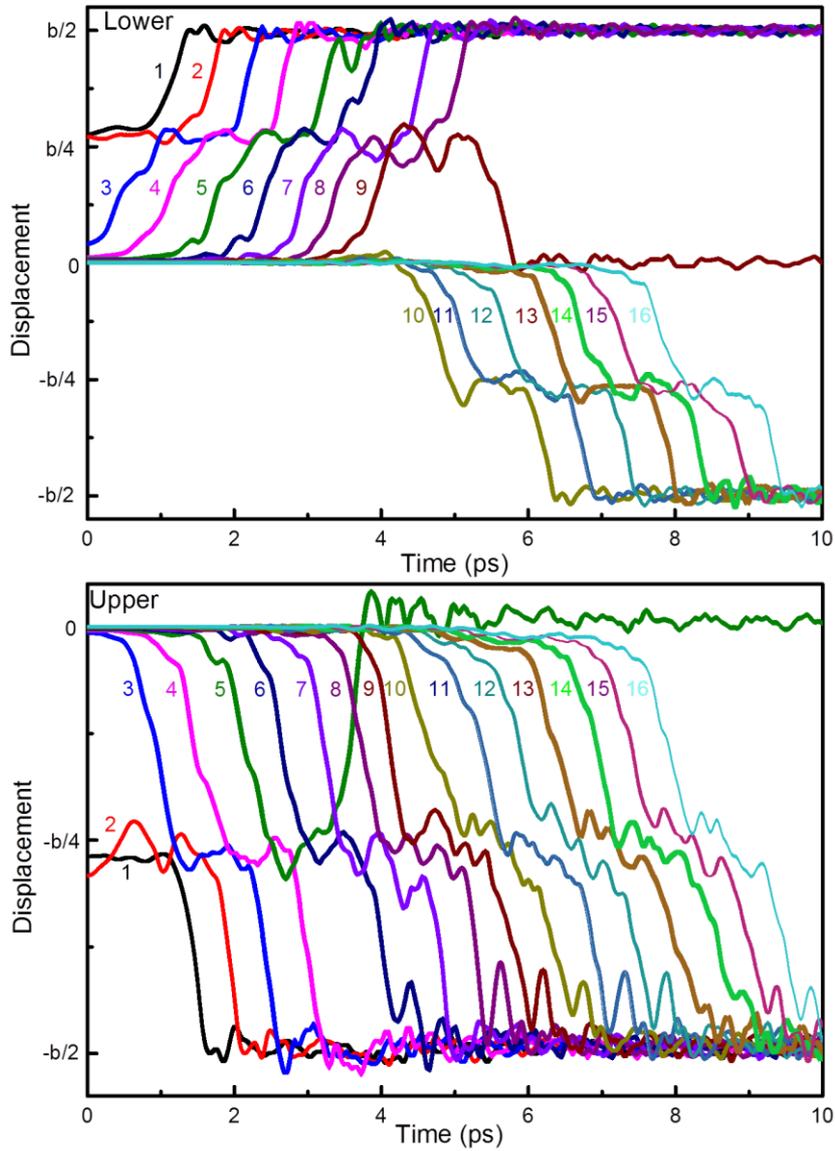

Figure 5 Atomic shifts along $b$ in both the lower and upper layer of Core-II subjected to a negative shear stress of 470 MPa, as a function of time. Movements in the other two directions are minor and irrelevant and thus not shown. The shifts for all atoms in the core that haven't completed a full shift of $b/2$, labeled from "1" to "16" are recorded. Reversal of upward shift to downward shift in the lower layer starts from atom No. 10.

It is clearly seen in Fig. 5 that, all atoms stayed at the $M$-site with a shift of $b/4$ transiently before they complete a full shift of $b/2$, confirming the stability of $M$-site occupancy. Atoms No. 1~8 in the lower layer all experience regular upward shift of $+b/2$; atom No. 9 also experiences regular upward shift of $+b/4$ to the $M$-site first, and then starts to move back at $t=5$ $ps$ as atom No. 10 and more of its successors reverse their shift direction. Meanwhile, all atoms in the upper layer remain regular downward shift, except for an atom No. 5 that moves back at an earlier time of 3 $ps$. Hence, double cross-slip down of the entire dislocation by one atomic layer through the sudden reversal of shift direction, without formation of jog and vacancies, is confirmed to happen for Core-II. Actually, similar double cross-slip behaviors, including continuous double cross-slip up, down, and down first then up are all detected for the cores reported in Fig. 4 (except for the stuck Core-I) at stresses of 360~500 MPa. Given the fact that reversal of shift direction and double cross-slip are extensively observed for strictly ideal, infinite-long dislocations at relatively low stresses, only 20%~60% higher than the Peierls stress, it is reasonable to conclude that the reversal of shift direction and double cross-slip are intrinsic features for slip of screw Py-II $<c+a>$ dislocations.

In addition to the unusual reversal process, pattern of regular atomic motion during dislocation motion is also revealed in Fig. 5. When the Core-II is subjected to shear stress, atoms in the core are not shifting in regular order as expected, namely atom No. 1 shifts first, followed by No. 2 and 3 and so on. Instead, they all move in a particular alternate order. For example, in the lower layer it is atom No. 3 (blue line in Fig. 5) initially staying at a lattice site with zero shift (referred to as $L$-site hereafter) that starts to shift first. As atom No. 3 approaches the stable $M$-site, increasing the number of atoms staying at $M$-sites from 2 to 3, atom No. 1 (black line in Fig. 5) at the $M$-site starts to leave *M*-site and complete the other half shift of *b*/4, restoring the number of atoms staying at *M*-sites from 3 to 2 again. Accordingly, the Core-II also advances by $\sqrt{3}\,a$ after a transient transformation to Core-III, and continues to advance as the same alternate process repeats again and again. Hence the core configuration of a pyramidal dislocation in motion does not stay constant, but varies periodically. Most of the time it appears as Core-II, and Core-III and others can only be observed transiently. Exact the same alternate shift order happens in the upper layer for atom pairs of No. 3 and 1, No. 4 and 2, No. 5 and 3 and so on, with slight time lag compared to that in the lower layer. The particular alternate order of atomic shift during dislocation motion revealed here should be attributed to the existence of a stable transit site (*M*-site) along the shift path.

**4. Energy landscapes associated with atomic shifts**

Now that the reversal of shift direction during pyramidal slip is confirmed to be intrinsic, there ought to be an energetic plausibility allowing the reversal process to happen. We next propose a novel three-step method to examine the energy landscapes associated with atomic shifts to figure out why the unusual reversal is happening. If a dislocation is actually moving, the energy landscapes associated with atomic shifts are also dynamically varying, making the examination extremely difficult. Alternatively, a static state close enough to the state of motion would be a better choice. Hence, we first apply a negative shear stress of 240 MPa, only 20% lower than the Peierls stress, to an infinite-long screw Py-II <*c*+*a*> dislocation. The dislocation remains still and thus static energy landscapes are acquirable.

Second, we assume that the shifts in the core take place in a sequential manner, one after another, and simultaneous shifts are not considered. This assumption turns out to be very reasonable, as evidenced by the actual atomic shifts recorded in Fig. 5. Then the energy landscape associated with each single shift is simple, and is calculated by vertically shifting (along ***b***) one vertical pair of target atoms in the core at a time. The target pair is consisted of one atom from each array and is allowed to move freely in the other two directions during energy minimization. All other atoms are free to move in all directions, except for those fixed in the top and bottom layers to maintain the shear stress.

Third, we determine the shift sequence as below: for an initial configuration, the first shift in the lower or upper layer happens to the pair that has lowest energy barriers amongst all possible candidate pairs; then the minimum-energy configuration during the first shift is chosen, and the second shift also happens to the pair with lowest energy barriers and so on. The novel three-step method proposed here to determine the energy landscapes associated with change or propagation of complex configurations can also be applied to other lattice defects without loss of generality.

Taking Core-II as an example, in the lower layer the first shift might happen to any of the four vertical pairs containing atom No. 1, 2, 3 and 4. Then the energy change as a function of shift displacement for each pair is calculated and the pair containing atom No. 3 turns out to have the lowest energy barriers and will shift first, in accordance with the displacement records in Fig. 5. The energy change for the pair containing atom No. 3 shifting by $b/2$ from an $L$-site through an $M$-site to another lattice site (referred to as $L'$-site hereafter) at 240 MPa is shown in Fig. 6A. The energy change at equilibrium state is also calculated for comparison. It should be pointed out that the strict periodicity for the vertical shift is $b$; however, the cyan atom No. 3 can take its neighboring pink atom's position with a vertical shift of $b/2$ and slight off-plane and [01-10] shifts, to restore periodicity, as shown in the left column of Core-II in Fig. 4. Hence the actual periodicity becomes $b/2$ here, and $L$-site and $L'$-site become identical. Consequently, the atomic configuration of Core-II becomes geometrically symmetric with respect to [01-10] direction (horizontal axis in Fig. 4) in both the lower and upper layer, and it's likely that upward and downward shifts of $L$-site atoms are geometrically equivalent.

It is seen from Fig. 6A that, $L$-site (zero shift) and $L'$-site (shift of $b/2$) are indeed energetically identical, and between them there exist multiple minima corresponding to stable configurations at the equilibrium state, explaining the polymorphism in the core configuration. In spite of the geometrical symmetry in the lower and upper layer, the energy landscape for upward and downward shifts at the equilibrium state is asymmetric: the upward shift of atom No. 3 from *L*-site towards *L'*-site turns out to be more difficult than the downward shift from *L'*-site towards *L*-site as indicated by their different energy barriers of 0.06 verses 0.008 eV. The asymmetry in the energy landscape must be attributed to the contributions from farther layers. Interestingly, as the negative shear stress increases to 240 MPa, the energy barrier for upward shift decreases, as expected, to 0.003 eV, indicating upward shift is about to happen; while the barrier for downward shift unexpectedly stays at 0.009 eV, meaning downward shift is also energetically favorable. The fine minima in the energy landscape also vanish and a single minimum with very low energy of ~ −0.25 eV develops at the *M*-site. Thus, the asymmetric energy landscape for the geometrically symmetric Core-II gets corrected by the negative shear stress and becomes symmetric again. The symmetry in the energy landscape developed around the *L*-site actually physically explains why the unusual reversal of shift direction intrinsically happens during dislocation motion: the *L*-site in dislocation core becomes bifurcation point as the negative stress approaches its critical value of ~300 MPa.

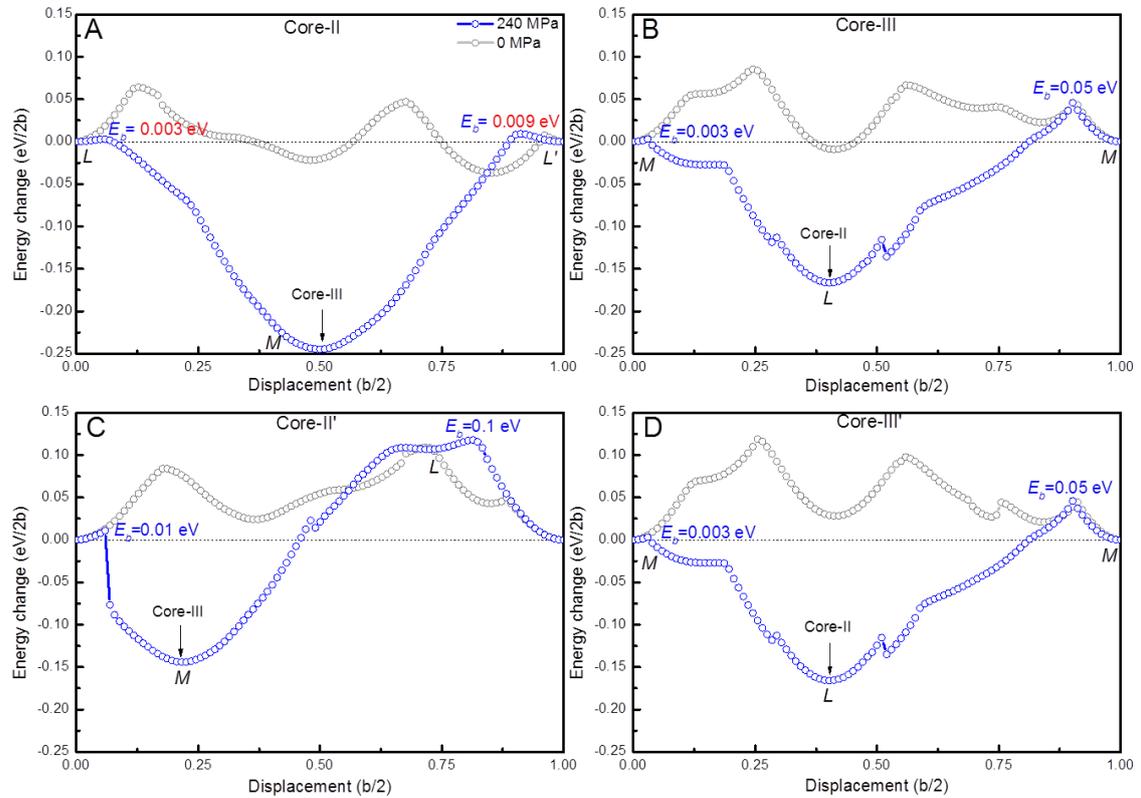

Figure 6 Energy landscapes associated with four sequential atomic shifts that advance the Core-II by a minimum period of √3 a, at both 0 and 240 MPa. Multiple minima appear in all cases at 0 MPa, confirming the polymorphism. The asymmetry is found around the *L*-site during the first shift (atom No. 3) in Core-II at 0 MPa (gray circles in Fig. 6A), and a bifurcation point emerges as the symmetry gets restored around the *L*-site (blue circles in Fig. 6A) due to the negative stress of 240 MPa. Shift in either upward or downward direction then becomes possible, explaining why the reversal happens intrinsically.

After the first shift by *b*/2 to reach the *M*-site, the Core-II correspondingly transforms to Core-III. Then the candidate pairs in Core-III containing atom No. 1, 2, 3 and 4 are examined again, and the second shift turns out to happen for the pair containing atom No. 1 at the *M*-site, in accordance with the displacement records in Fig. 5 again. The energy change for the pair containing atom No. 1 shifting from *M*-site through an *L*-site to another *M*-site is shown in Fig. 6B. Similarly, multiple minima exist at the equilibrium state. As the shear stress increases to 240 MPa, the energy barrier for upward shift decreases to 0.003 eV, while the barrier for downward shift stays at a high level of 0.05 eV, indicating that only normal upward shift is favorable.

The second shift by ~*b*/2 to reach an *L*-site transforms Core-III back to Core-II with an advance of √3 *a* in [01-10] direction in the lower layer. Then we continue to examine the candidate pairs in the upper layer of the Core-II with an advance of √3 *a*. The configuration of the upper layer is close to Core-II' shown in Fig. 4, with 2 atoms at *M*-sites and an additional atom with a shift between 0~*b*/4, and the third shift turns out to happen for the pair with a shift between 0~*b*/4. The energy change for the pair shifting through an *M*-site in the upper layer is shown in Fig. 6C. Similarly, multiple minima exist at the equilibrium state. As the shear stress increases to 240 MPa, the energy barrier for normal downward shift decreases to 0.01 eV, while the barrier for upward shift stays at a high level of 0.1 eV, indicating that in the upper layer only normal downward shift is favorable.

The third shift by ~*b*/8 to reach an *M*-site transforms the core to Core-III' shown in Fig. 4, namely a core with 3 atoms at *M*-sites in the upper layer. Similar to Core-III with 3 atoms at *M*-sites in the lower layer, the fourth shift happens for the last pair at the *M*-site, and its energy landscape appears to be almost identical to the second shift in the lower layer, as shown in Fig. 6D. After four sequential shifts, two in each layer, the initial Core-II fulfills a complete advance of the whole dislocation core by a minimum period of $\sqrt{3}\ a$ in both layers, and restores its configuration.

Amongst the four shifts involved in the unit advance of the dislocation core, only the first shift starting from an *L*-site exhibits symmetry and is a bifurcation point that can go either downwards or upwards, a root cause for all the unusual processes revealed in the present study. The other three shifts are all energetically asymmetric upon directional stress, and can only go unidirectionally in a conventional manner.

Higher stresses of 250~290 MPa are also applied to Core-II and similar results are obtained, only that instabilities, seen as the spikes in Figs. 6B-D, start to develop as the stress approaches the Peierls stress and the core is about to set off. Starting with a different core also ends up with the same results, since these transient configurations during dislocation motion are already covered when examining a full-period motion of the Core-II.

**Discussion**

The energy landscapes associated with atomic shifts in the core provided here reveal the symmetry and bifurcation developing at the *L*-sites as shear stress increases, which are the root cause for the intrinsic reversal of shift direction observed during slip of screw Py-II <*c+a*> dislocations. The reversal of shift direction then leads to double cross-slip, and the inevitable asynchrony of double cross-slip leads to formation of jogs and motion of the jogs generates massive vacancy clusters.

The slip of pyramidal dislocations can generate point defects throughout the sample. Such massive point defects spreading out in the sample then would have profound impact on the mechanical behavior of Mg. First, the massive point defects on Py-II planes could hinder subsequent dislocation motion and lead to significant strain hardening, which is indeed observed for Mg [1-3]. Second, the vacancy clusters themselves are permanent damages in the crystal. As plasticity proceeds, the accumulation of damage gets severer and eventually leads to pre-mature fracture and brittleness. Third, double cross-slip and generating jogs and defects dissipate dislocation energy and slow down dislocation velocity, and thus significantly reduce the dynamic performance of Mg. Quantification of such behaviors of jogged dislocations and their contribution to mechanical behavior in bcc metals and other materials have also been investigated by theoretical [31, 32] and kinetic Monte Carlo models [33, 34].

In a word, the massive point defects can be a main contributor to high hardening, brittleness and poor dynamic performance of Mg. Given the dominance of screw slip and the intrinsic nature of the reversal of shift direction, double cross-slip, jog formation and point defect generation during pyramidal screw slip, even if we are able to activate sufficient amount of glissile pyramidal <c+a> dislocations, massive damage generation is still inevitable and improved property remains unachievable. Hence sufficient amount of pyramidal dislocations could still be insufficient to effectively accommodate plasticity. In order to essentially achieve improved property, the strategy of stabilizing Py-II screw slip by appropriate alloying comes with first priority and is fundamental and of vital importance.

Recently there also have been efforts attempting to screen appropriate alloying elements for enhancement of pyramidal dislocation formation [35] using DFT calculations. However, for pyramidal dislocation slip, the actual effects of alloying elements in the absence of a concrete dislocation core like the ones presented in Fig. 4, remain unknown. Hereby we would like to draw the attention of the community and emphasize the importance of a concrete dislocation core in such quests, especially when the behavior of <c+a> dislocations in hcp Mg has far exceeded our conventional understanding and expectations, as demonstrated in the present study. With the concrete cores provided here and the energy landscape examination method proposed here, effects of alloying remedy on screw <c+a> dislocation motion can then be practically evaluated, in the aim of stabilizing dislocation motion to avoid point defects and to achieve enhanced plasticity and strength.

The study presented here reveals the unusual massive point defect generation accompanying pyramidal slip in hcp metal Mg, its double cross-slip mechanism and bifurcation in atomic shifts as a root cause. The intrinsic instability in atomic shifts during pyramidal slip discovered in Mg could also be a common feature for a wider range of crystalline materials with hcp lattice or other lattice with low symmetry and large Burgers vectors, such as ceramics, and the specific findings made in Mg provide an opportunity to better understand the complicated slip behavior and instabilities in these brittle materials.

It should also be noted that so far direct experimental observation of point defects is still very challenging in materials science, even with advanced quantitative scanning transmission electron microscopy (STEM) techniques [36]. Randomly distributed vacancies just don't produce contrast between different columns of atoms in TEM images. Even so, there are still chances that direct observation becomes possible when there is an ordering and continuous form of vacancies [37], just as the continuous arrays of vacancies we showed in Figs. 1 and 3. Careful TEM examinations of continuous arrays of vacancies in deformed Mg samples in the future are thus expected to further verify the observed double cross-slip and point defect generation here.

**Data Availability**

The data that support the findings of this study are available from the corresponding author upon reasonable request.

**Author contributions**: Y.T. contributes solely to the research.

**Competing financial interests**: The author declares no competing financial interests or non-financial interests.

**Materials & Correspondence:** Please send all correspondence and materials requests to Yizhe Tang (yzhe.tang@gmail.com).